\documentclass[12pt]{iopart}

\usepackage[dvips]{graphicx}

\begin{document}

\title{General Relativistic Galaxy Rotation Curves: Implications for Dark Matter Distribution}

\author{Dylan Menzies and Grant J. Mathews}

\address{Department of Physics, Center for Astrophysics, University of Notre Dame, Notre Dame, IN 46556  email:  dmenzies@nd.edu,  gmathews@nd.edu}

\begin{abstract}
It has recently been suggested  that observed galaxy rotation curves can be accounted for by general relativity without recourse to dark-matter halos.  Good fits have been produced to observed galatic rotation curves using this model.  We show that the implied total mass is infinite, adding to the evidence opposing the hypothesis.
\end{abstract}

\maketitle

\section{Introduction}
Understanding the nature and origin of the apparent dark matter responsible for the bulk  of the gravitational mass of galaxies and clusters is one of the most important challenges in modern cosmology.
Although many candidates have been proposed for the required weakly interacting cold dark matter \cite{Baltz04},
to date there has been no convincing detection of a suitable particle candidate.  For this reason there is considerable interest in a recent proposal \cite{CT} that accurate modeling of galactic dynamics using general relativity is consistent with observed flat rotation curves and luminous mass without the need for additional dark matter. In this scenario, the authors assume a cylindrical stationary space-time containing pressureless dust and solve the weak-field Einstein equation to first order in the gravitational constant $G$ to obtain the associated velocity distribution as described below.

At first glance, this would seem to be a compelling solution to the dark matter problem.
However, objections have already been raised by several authors, on a various grounds.  These include   the conditions at $z=0$ \cite{korz,VL}, general approximation principles \cite{garf}, and specific analysis of the approximations used \cite{cross}. The original authors have responded to some of these criticisms \cite{response}.   In this paper, we suggest yet another possible problem with this interpretation of the dark-matter problem.  We reanalyze the galactic rotation curve as in \cite{CT} and find that
the model, as applied, contains an implied large (infinite) mass distribution extending beyond the
observed galactic disk, and hence at best has simply moved the dark-matter to the outer Galaxy, and
more likely, the model is ruled out on energetic grounds.

\section{The Cooperstock and Tieu Model}
Our goal is to analyze the implied mass distribution in the Cooperstock \& Tieu \cite{CT}  model.
In order to deduce this we first summarize their approach.
In their analysis of  rotation curves, the galaxy to be described is treated as
a uniformly rotating axisymmetric system containing  pressureless fluid.
The general stationary axially symmetric
metric is chosen of  the form
\begin{equation}
	ds^2 =	-e^{\nu-w}( udz^2+dr^2)
		-r^2 e^{-w} d\phi^2+e^w(cdt-Nd\phi)^2~~,
\label{Eq1}
\end{equation}
where $u$, $\nu$, $w$ and $N$ are functions of cylindrical polar
coordinates $r$, $z$, and the metric coefficient
$u$ is taken to be unity.  In a co-moving frame the
the fluid four-velocity is just
\begin{equation}
	U^i = {\delta}_0^i ~~.
\label{Eq2}
\end{equation}
 Making  a local
($r$, $z$ held fixed) transformation to
\begin{equation}
	\bar{\phi} = \phi + \omega(r,z) \,t~~,
\label{Eq3}
\end{equation}
 diagonalizes the metric and,  in the weak field limit, leads to simple approximations for
  the local angular velocity $\omega$ and the tangential
velocity $V$:
\begin{equation}
	\omega
	= \frac{Nce^w}{r^2e^{-w}-N^2e^w}
	   \approx  \frac{Nc}{r^2}~~,  \\
	V =\omega r~~.
\label{Eq4}
\end{equation}

The Einstein field equations to first order in the gravitational constant $G$
 are easily combined to yield a Poisson-like equation for $\omega$.
\begin{equation}
	\nabla^2 w +\frac{N_r^2+N_z^2}{r^2}=\frac{8\pi G\rho}{c^2}~~,
\end{equation}
where
\begin{equation}
	\nabla^2 w \equiv w_{rr} + w_{zz} + \frac{w_r}{r}~~.
\end{equation}

The geodesic
equation
\begin{equation}
	\frac{dU^i}{ds} +\Gamma^i_{kl}U^k U^l=0
\end{equation}
together with the adopted metric and four-velocity
 (\ref{Eq1}) and (\ref{Eq2}) simply becomes
\begin{equation}
	w_r=w_z=0~~,
\end{equation}
and hence
\begin{equation}
	\nabla^2 w=0~~,
\end{equation}
within the fluid.  The interior equations for $\rho$ and $N$ then become:
\begin{equation}
	N_{rr} + N_{zz} - \frac{N_r}{r} =0
\label{Eq9a}
\end{equation}
\begin{equation}
		\frac{N_r^2 + N_z^2}{r^2} = \frac{8{\pi}G\rho}{c^2}~~.
\label{Eq9b}
\end{equation}
Eq. (\ref{Eq9a})
can be expressed as
\begin{equation}
	\nabla^2\Phi =0
\label{Eq10}
\end{equation}
where
\begin{equation}
	\Phi \equiv \int\frac{N}{r}dr~~.
\label{Eq10a}
\end{equation}
Hence,  flat-space harmonic functions $\Phi$ become  the generators of the
axially symmetric stationary pressure-free weak fields.
 Using (\ref{Eq4}) and (\ref{Eq10a}),  the
tangential velocity distribution becomes
\begin{eqnarray}
	V &=& c\frac{N}{r} \\
	&=& c\frac{\partial {\Phi}}{\partial{r}}~~.
\end{eqnarray}
Since the field equation for $\rho$ is non-linear, the galactic rotation is deduced
by first finding the required generating
potential $\Phi$ and from this the appropriate function $N$
for the galaxy is deduced. Then Eq.~(\ref{Eq9b})
yields the density distribution.

Each galaxy analyzed requires its own composing elements to
build the generating potential. A separation of variables yields a solution for $\Phi$ in terms of Bessel functions $J_n(kr)$
\begin{equation}
	\Phi = Ce^{-k\mid z \mid}J_0(kr)
\end{equation}
where $C$
is an arbitrary constant. The general solution is expressed as a linear superposition
\begin{equation}
	\Phi = \sum_{n}C_ne^{-k_n |z|}J_0(k_nr)
\label{Eq13}
\end{equation}
with $n \sim 10$ chosen appropriately for the desired level of accuracy.
From (\ref{Eq13}) and (\ref{Eq4}), the tangential velocity is
\begin{equation}
	V= -c\sum_{n} k_n C_n e^{-k_n |z|}J_1(k_nr)~~.
\end{equation}

This expression was used in \cite{CT} to fit the rotation curve for the Milky Way.
In this paper we analyze this fit and show that any fit implies a large mass at larger radii.
Hence, this approach does not solve the dark matter problem, but has merely moved to
dark matter to large radii. Worse, we now have an excess mass problem!

\section{Interpretation of the Model Fits}

\begin{figure}
\begin{center}
\includegraphics[width=8cm]{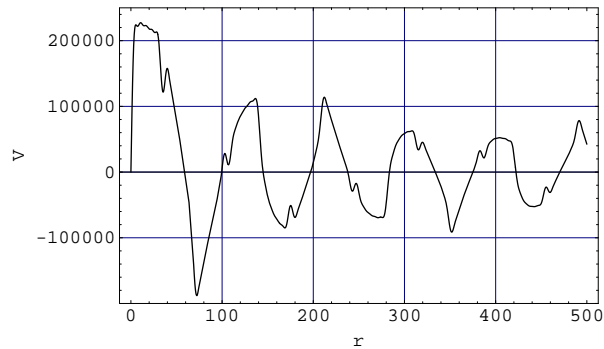}
\caption{\label{fig:vel500} Velocity $V \ (m/s)$ as a function of radius $r \ (kpc)$}
\end{center}
\end{figure}

\begin{figure}
\begin{center}
\includegraphics[width=10cm]{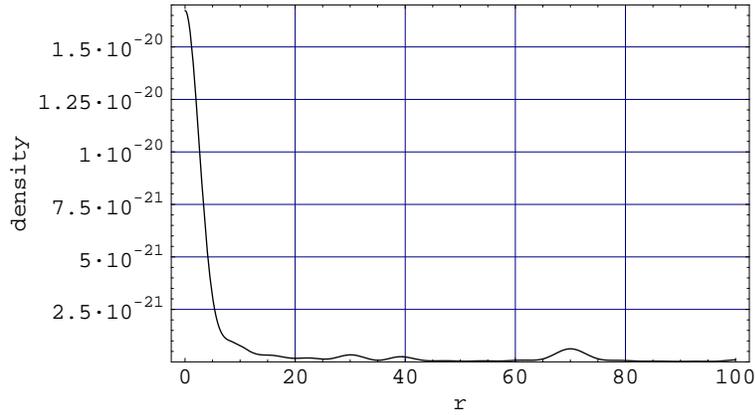}
\caption{\label{fig:density100} Density $\rho \ (kg m^{-3})$ as a function of radius $r \ (kpc)$}
\end{center}
\end{figure}

Figure \ref{fig:vel500} reproduces the velocity curve fit for the Milky Way given in \cite{CT}, but extended to larger radius. The large oscillations in velocity although alarming, are rendered harmless if the density is sufficiently low. Figure \ref{fig:density100} shows the extended density profile. It would be tempting to assume from this  that the density could be cutoff at the limits of the observed galaxy. However, we show that for any velocity fit the total mass is infinite, in contrast  to the original claim of finite integrated mass. In fact, smoothed out,  the enclosed mass rises linearly with radius.

\section{Integrating the mass}
\begin{figure}
\begin{center}
\includegraphics[width=10cm]{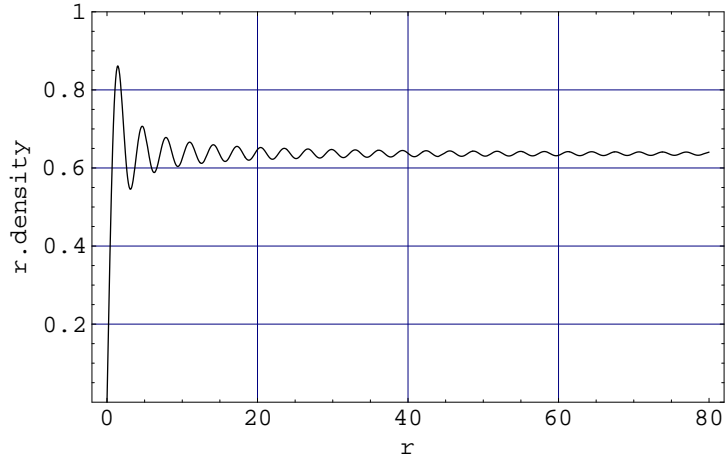}
\caption{\label{fig:rdensityk} $r \rho$ as a function of radius $r \ (kpc)$, $z = 0$}
\end{center}
\end{figure}

To obtain the total implied galactic mass, consider a model with a single value of $k$,
\begin{equation}
N = r e^{-kz} J_1(kr) ~~.
\end{equation}
Eq (\ref{Eq9b}) written for density is
\begin{equation}
\label{eq:density}
\rho = \frac{c^2}{8\pi G}\left(\frac{N_r^2+N_z^2}{r^2}\right) ~~.
\end{equation}
In this case
\begin{equation}
\rho = e^{-2kz} (J_0(kr)^2 + J_1(kr)^2)~~ .
\end{equation}
We note $r \rho \rightarrow({2}/{\pi} ) \exp{\{-2z\}}$ as $r \rightarrow \infty$, illustrated in Figure \ref{fig:rdensityk}.
The mass element is
\begin{equation}
dm = \rho dr dz \, r d\theta~~.
\end{equation}
So the mass enclosed within  cylindrical coordinate r is
\begin{eqnarray}
\label{eq:M}
M(r) &=& 2 \pi \int dr dz \, r\rho \\
 &\rightarrow& \biggl[2 \pi \biggl(\frac{2}{\pi}\biggr) \int dz e^{-kz} \biggr] r~~ .
\end{eqnarray}
In this case $M(r)$ clearly rises linearly.

\begin{figure}
\begin{center}
\includegraphics[width=10cm]{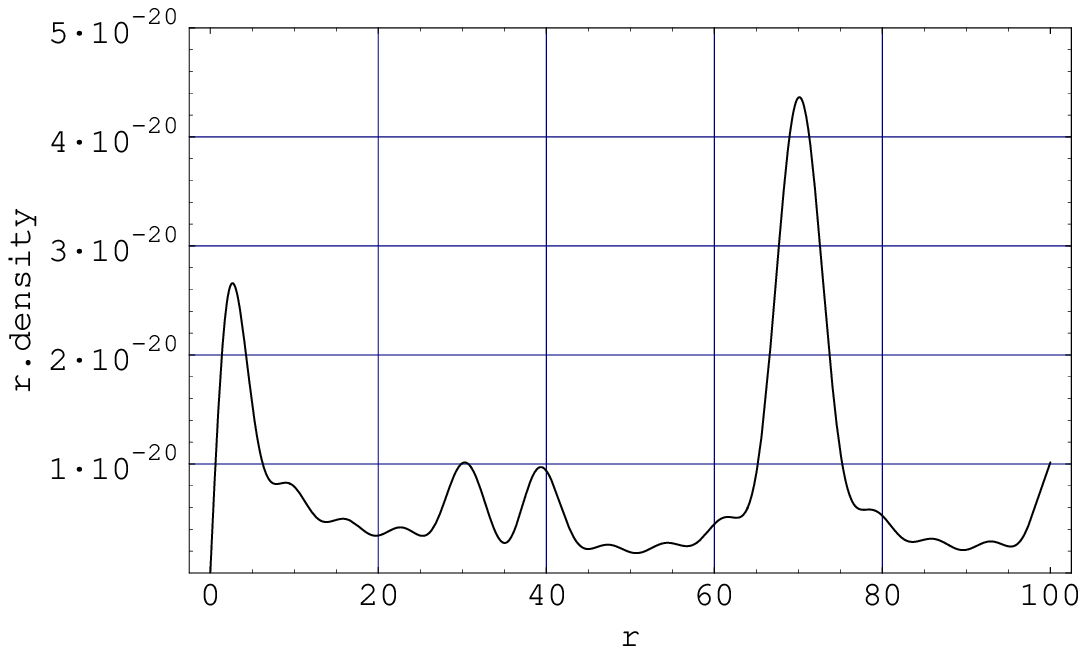}
\caption{\label{fig:rdensity100} $r \rho$ as a function of radius $r \ (kpc)$, $z = 0$}
\end{center}
\end{figure}

\begin{figure}
\begin{center}
\includegraphics[width=10cm]{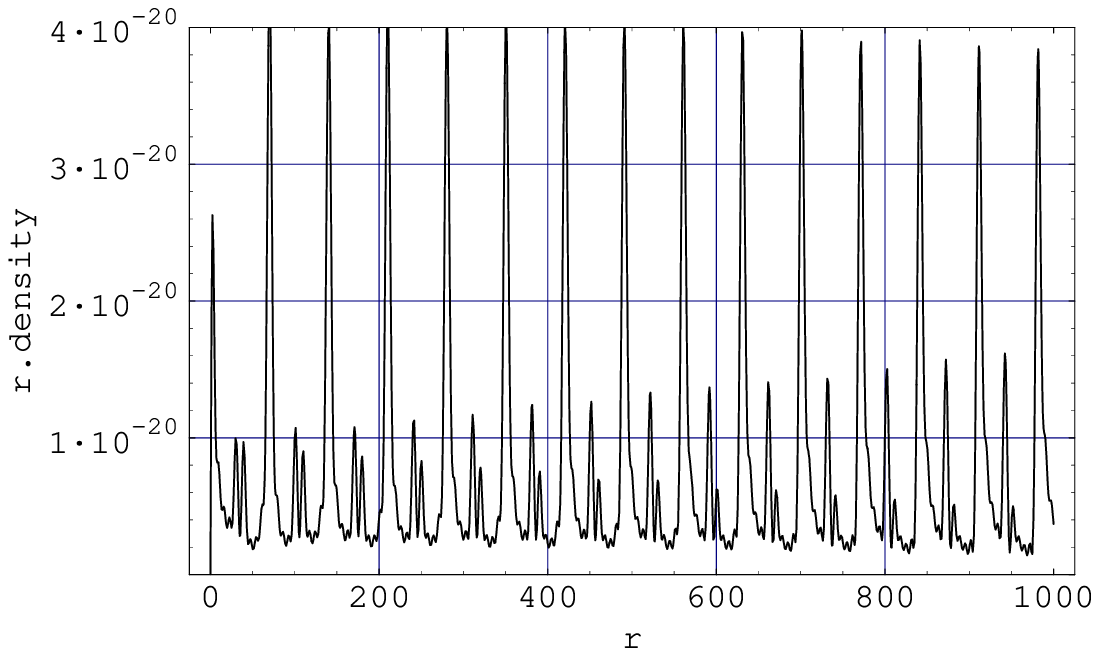}
\caption{\label{fig:rdensity1000} $r \rho$ as a function of radius $r \ (kpc)$, $z = 0$}
\end{center}
\end{figure}

Now we look at the general case. Taking our cue from Eq.(\ref{eq:M}), $r \rho$ is plotted against $r$ for the Milky Way model using two different scales of extended range, shown in Figures \ref{fig:rdensity100}, \ref{fig:rdensity1000}.
It appears that $r \rho$ oscillates around a positive average value. If the $z$ dependence were independent of $r$ like that in the single $k$ case, this would imply that $M(r)$ tends to rise linearly in the limit of large $r$. The following analysis shows this is the case, even though the $z$ dependence is not simple.

If we can find $\alpha, \beta>0$ such that for large enough $r$
\begin{equation}
\frac{\int dz r\rho(r,z)}{r\rho(r,0)} > \alpha
\end{equation}
and
\begin{equation}
\int_0^r r\rho(r,0) > \beta r
\end{equation}
then $M(r)> \alpha\beta r$ for large enough $r$. To find $\alpha$, for fixed $r$ define
\begin{equation}
f(z) = \frac{r\rho(r,z)}{r\rho(r,0)} .
\end{equation}
From (\ref{eq:density}),
\begin{equation}
f(z) = (\sum_k a_k e^{-k z})^2 + (\sum_k b_k e^{-k z})^2 ,
\end{equation}
with coefficients $a_k, b_k$ determined by the model. Note $f(0) = 1$. Define $\alpha$ as the minimum over all sets ${a_k, b_k}$, in any model, of the integral,
\begin{equation}
\alpha = \min_{\{a_k, b_k\}} \int dz f(z) .
\end{equation}
$\alpha > 0$ since otherwise $f(z) \ge 0$ and continuity of $f(z)$ imply $f(0)=0$, a contradiction.
\\
\\
To find $\beta$ first find the form of the density for $z=0$ using (\ref{eq:density}),
\begin{equation}
\rho(r,0) = \left[ \left(\sum_k \left( r a_k J_0(kr) + c_k J_1(kr)\right)\right)^2 + \left(\sum_k r b_k J_1(kr)\right)^2 \right]/r^2 ,
\end{equation}

with coefficients determined by the model. So as $r \rightarrow \infty$
\begin{eqnarray}
r\rho &\rightarrow& r\left[ \left( \sum_k a_k J_0(kr) \right) ^2 + \left(\sum_k b_k J_1(kr)\right)^2 \right] \\
      &\rightarrow& \left( \sum_k  a_k \sin{kr} \right)^2 + \left( \sum_k  b_k \cos{kr} \right)^2 ,
\end{eqnarray}
with a re-scaling of the coefficients. So the integral
\begin{eqnarray}
\int dr r\rho = \sum_k \int_0^r dr (a_k \sin{kr})^2 + (b_k \cos{kr})^2 \\
 +  \sum_{k_1 \ne k_2} \int_0^r dr \left( a_{k_1}a_{k_2} \sin{k_1 r} \sin{k_2 r} + b_{k_1}b_{k_2} \cos{k_1 r} \cos{k_2 r} \right) \nonumber
\end{eqnarray}
The first sum consists of a piece linear in $r$ plus an oscillating piece, while the second sum is purely oscillating. The oscillations are all bounded, so overall the required $\beta>0$ exists, and $M(r)$ behaves as predicted.

\section {Conclusion}
Despite the close fit to observed velocity curves over the galatic range, we find that the total enclosed mass in the model of Cooperstock and Tieu rises as a linearly increasing function of radius. This indicates that the model has not removed the dark matter problem, but has merely moved the mass to large radius.  Moreover, the implied infinite mass, and large velocity fields suggest that this interpretation is probably not realistic.

\vskip .2 in
Discussions with Bill McGlinn are gratefully acknowledged. Work at the University of Notre Dame supported by the U.S. Department of Energy under research grant DE-FG02-95-ER40934.


\section*{References}
\begin{thebibliography}{99}

\bibitem{Ostriker03} Ostriker, J. P. and Steinhardt, P. 2003, Science, 300,  1909
\bibitem{Baltz04} Baltz, E. A. 2004, astro-ph/0412170
\bibitem{CT} Cooperstock, F.I. and Tieu, S. 2005, astro-ph/0507619
\bibitem{korz} Korzynski, M. 2005, astro-ph/0508377
\bibitem{VL} Vogt, B. and Letelier, 2005 P.S., astro-ph/0510750
\bibitem{garf} Garfinkle, D. 2005,  Class.Quant.Grav. 23 (2006) 1391, gr-qc/0511082
\bibitem{cross} Cross, D.J. 2006, astro-ph/0601191
\bibitem{response} Cooperstock, F.I. and Tieu, S. 2005, astro-ph/0512048

\end{thebibliography}
\end{document}